\documentclass[aps,amsmath,amssymb,prd,floatfix,preprint,superscriptaddress,nofootinbib,12pt]{JHEP3}
\usepackage{epsfig}
\usepackage{amsmath}
\usepackage{amssymb,amsfonts}
\usepackage{comment}
\usepackage{graphicx}
\usepackage{latexsym}
\usepackage{slashed}
\usepackage{xcolor}
\usepackage{subfigure}
\usepackage{subfig}
\let\normalcolor\relax

\makeatletter
\def\@fpheader{\relax}
\makeatother

\relax
\renewcommand{\theequation}{\arabic{section}.\arabic{equation}}
\def\be{\begin{equation}}
\def\ee{\end{equation}}
\def\bs{\begin{subequations}}
\def\es{\end{subequations}}
\newcommand{\een}{\end{subequations}}
\newcommand{\ben}{\begin{subequations}}
\newcommand{\beq}{\begin{eqalignno}}
\newcommand{\eeq}{\end{eqalignno}}
\def\g{\gamma}

\def\m{\mu}
\def\n{\nu}

\def\sp{\;\;\;,\;\;\;}

\newcommand\fverb{\setbox\pippobox=\hbox\bgroup\verb}
\newcommand\fverbdo{\egroup\medskip\noindent%
                        \fbox{\unhbox\pippobox}\ }
\newcommand\fverbit{\egroup\item[\fbox{\unhbox\pippobox}]}
\newbox\pippobox

\def\hri#1#2{\href{http://arxiv.org/abs/#1}{[ArXiv:#1]#2}}
\def\hre#1#2{\href{http://arxiv.org/abs/#1/#2}{[ArXiv:#1/#2]}}

\def\beq{\begin{equation}}
\def\eeq{\end{equation}}

\def\4R{{{}^{(4)}R}}
\def\H{{\cal H}}
\def\K5{{\kappa}}
\def\K52{{\kappa^2}}

\newcommand{\half}{\frac{1}{2}}

\newcommand{\pa}{a^{\prime}}

\newcommand{\ba}{\begin{aligned}}
\newcommand{\ea}{\end{aligned}}

\relax
\renewcommand{\theequation}{\arabic{section}.\arabic{equation}}

\def\be{\begin{equation}}
\def\ee{\end{equation}}

\newcommand{\ha}{{1 \over 2}}

\newcommand{\de}{\partial}
\newcommand{\bea}{\begin{eqnarray}}
\newcommand{\eea}{\end{eqnarray}}

\def\hri#1#2{\href{http://arxiv.org/abs/#1}{[ArXiv:#1]#2}}
\def\hre#1#2{\href{http://arxiv.org/abs/#1/#2}{[ArXiv:#1/#2]}}

\newbox\pippobox

\def\II{\relax{\rm I\kern-.18em I}}

\def\m{\mu}
\def\n{\nu}

\def\g{\gamma}

\def\pa{\partial}

\def\sp{\;\;\;,\;\;\;}

\def\f{\varphi}

\title{Brane Cosmology and the self-tuning of the cosmological constant in the presence of bulk black holes}
\author{P. Betzios$^{\flat}$,  O. Papadoulaki$^*$\\
~\\
$^\flat$ \href{http://hep.physics.uoc.gr}{Crete Center for Theoretical Physics}, Institute for Theoretical and Computational Physics,
Department of Physics,  P.O. Box 2208,\\
University of Crete, 70013, Heraklion, Greece
~\\

$^*$ \href{https://www.ictp.it/}{International Centre for Theoretical Physics} \\
Strada Costiera 11, Trieste 34151 Italy.
}
\preprint{CCTP-2020-2\\ITCP-IPP-2020/2}

\abstract{Motivated by the holographic self-tuning proposal of the cosmological constant, we generalize and study the cosmology of brane-worlds embedded in a higher-dimensional bulk black hole geometry. We describe the equations and matching conditions in the case of flat, spherical and hyperbolic slicing of the bulk geometry and find the conditions for the existence of a static solution. We solve the equations that govern dynamical geometries in the probe brane limit and we describe in detail the resulting brane-world cosmologies. Of particular interest are the properties of solutions when the brane-world approaches the black hole horizon. In this case the geometry induced on the brane is that of de Sitter, whose entropy and temperature is related to those of the higher dimensional bulk black hole.}

\keywords{Braneworld, Cosmology, de Sitter, Holography}
\begin{document}

\section{Introduction, Results and Outlook}\label{intro}

One of the biggest and yet unsettled problem in theoretical physics is the explanation of the observed very small positive cosmological constant of our expanding universe or equivalently the explanation of an un-natural small vacuum energy. This is the so called cosmological constant problem  \cite{Weinberg:1988cp,Padilla:2015aaa,Burgess}. There are two ways to approach it from a low energy point of view, one is using the properties of effective quantum field theory (EQFT) and the other is using the framework of general relativity (GR).

In more detail using the EQFT paradigm, one can describe with much success the standard model (SM) of observed particles in the absence of gravity. If we use EQFT to compute the energy density of the vacuum, it receives large renormalisations originating from UV modes. Therefore unless an extreme fine-tuning of parameters takes place, one expects a huge energy density from such calculations. Simultaneously the vacuum energy density sources the gravitational fields that are responsible for the curvature of the universe thus it affects the macroscopic or classical (IR) physics and the evolution of the observed universe today. This is in stark conflict with the currently observed very small value of the cosmological constant.

There are various proposals for the resolution of the cosmological constant problem in the literature. In this paper we will focus on the self- tuning\footnote{By self-tuning of the cosmological constant (CC) we define any model that allows the relaxation of the CC into a smaller value dynamically, by allowing for extra degrees of freedom to backreact and exchange energy with the SM fields.} mechanism proposed in \cite{selftuning} that involves the use of the  brane-world scenario and holography\footnote{This mechanism was proposed in \cite{ArkaniHamed}, \cite{Kachru}. These efforts however did not manage to produce regular self-tuning solutions.}. In such a framework, the simplest model that encapsulates the relevant physics, is that of a five dimensional bulk space time that is a solution to an Einstein scalar theory. The geometry in the UV asymptotes to that of an $AdS_{5}$ space time. In this geometry we have embedded a four dimensional brane that corresponds to the observed universe with the standard model fields. Employing the holographic principle, the five dimensional bulk gravity theory is dual to a strongly coupled four dimensional quantum field theory that interacts with the four dimensional weakly coupled theory of the standard model brane \cite{rs, ArkaniHamed,  RZ, APR, smgrav}. The self-tuning of the cosmological constant in this framework is achieved by the coupling of the two sectors: the one that lives on the brane (which we will colloquially call the SM) and the one that is dual to the five dimensional bulk gravitational theory (which we will denote as the ``\emph{hidden sector}"~\cite{bbkn}). This interaction when we work in mass scales below that of any messenger fields that mediate interactions between the two sectors, can be described by introducing effective brane potentials in front of the kinetic terms of the fields that live on the standard model brane. In the present simple example, these potentials are functions of the scalar field of the five dimensional gravitational theory.

In the setup of \cite{selftuning}, one should impose Israel junction conditions at the radial point where the four dimensional brane is inserted and solve them together with the five dimensional bulk equations of motion. The authors then searched for stable solutions with zero cosmological constant induced on the standard model brane and found that such solutions exist and are stable under small perturbations, given some assumptions about the scalar potential induced on the brane. These solutions correspond to the Poincar\'e invariant vacua of the theory. Moreover, the brane stabilizes at some point $r_{0}$ in the bulk of the five dimensional space-time. The main novelty of the self-tuning construction of \cite{selftuning}, is demanding regularity in the IR of the bulk geometry as a basic principle\footnote{At finite temperature this will translate into a regularity condition at the bulk horizon.}, and searching for a consistent solution where the brane stabilizes somewhere in the bulk space time. This is done assuming general effective potentials that couple the bulk theory with the one that lives on the brane and the construction is then found to be able to give solutions that evade the issues (singular solution in the IR) present at \cite{ArkaniHamed, Kachru, csaki}\footnote{One might argue that demanding IR regularity is a point where fine-tuning is introduced into the construction, but in fact this is not only  a natural boundary condition, but also a dynamically required one from the point of view of the dual QFT. It is precisely regularity that fixes the vevs as a function of the couplings.}. Finally a mechanism analogous to that introduced in \cite{DGP}, results in an effective localisation (for some regime of scales) of the gravitational force in four dimensions, as perceived by observers that can perform experiments on the brane.

In \cite{Amariti:2019vfv} the authors studied time dependent curved solutions that are not static but rather move in the radial bulk direction. In contrast to \cite{dS}, the boundary of the five dimensional bulk theory is the one of \cite{selftuning}. Thus, these time-dependent solutions correspond to time-dependent states of the theory whose vacuum (static) solutions were found in \cite{selftuning}. The time dependence of the position of the brane  induces a time dependence on the brane metric, which leads to a cosmological evolution as perceived from the observers on the brane. Although the authors found the appropriate Israel junction conditions and bulk equations of motion, they could not solve them analytically and thus relied on a probe limit analysis where the brane does not backreact on the bulk geometry  (the bulk geometry remains static). In this limit, the time dependence enters only in the radial direction $r(\tau)$. Even taking such a limit, it is still possible to find cosmological solutions in agreement with the general principles in \cite{mirage}.

In the moving brane solutions, if the brane is moving towards the IR, the brane universe has a contracting FRW metric. On the other hand, moving towards the UV one finds an expanding FRW universe. Moreover, taking the UV limit where the bulk geometry asymptotes to $AdS$ one finds the induced geometry on the brane to be de Sitter with a Hubble parameter that  depends on the UV values of the potentials induced on the brane. In the IR on the other hand, there are several possibilities depending on how the bulk geometry ends. For example we can have acceptable forms of singularities or horizons.

In this paper, our main focus will be to extend these previous studies in the case where the bulk solution corresponds to a thermal state of the dual holographic QFT. In such a case there will be a regular bulk horizon cloacking the bulk  singularity. We believe this is one of the most interesting options physically, since it corresponds to geometries that describe the typical high energy states of the dual strongly coupled holographic theory, according to the eigenstate thermalisation hypothesis~\cite{Srednicki,DAlessio:2016rwt}. Thus, we expect our results to be quite robust and universal, especially in the limit where the brane approaches the regular horizon. In addition, having as a motivation a solution to the cosmological constant problem (for observers living on the brane), we would like to understand the properties of states where the total vacuum energy of the combined system is quite large. If the induced curvature on the brane can remain naturally low even in such a case, this is a strong indication that the mechanism proposed in \cite{selftuning} is quite robust.

In practice, we follow the steps of \cite{Amariti:2019vfv}, but this time we include regular black hole geometries in the five dimensional bulk. We first derive the full backreacting bulk equations of motion together with the Israel junction conditions across the brane, for three possible slicings of the bulk black hole geometry, namely flat, spherical and hyperbolic. These equations are coupled nonlinear ODE's and one can solve them only numerically.
Nevertheless existence of solutions for some parameter space can be established without having the precise explicit solutions.

In addition if we assume that the bulk is dual to a large-$N$ strongly coupled gauge theory, it is also natural to expect that the backreaction of the small number of SM degrees of freedom can be naturally kept low and hence there is a good motivation to further analyse a probe brane limit. In this limit since the only time dependence comes from the radial position of the brane, the action that governs the dynamics on the brane reduces to that of a quantum mechanical system. Moreover, the induced metric on the brane is of the FRW type and is controlled  by the blackening factor, the scale factor of the bulk geometry and the position of the brane in the radial direction.

The next step is to study the induced geometries on the brane in the IR and the UV limit. The UV asymptotics in the case of flat slicing are the same as in \cite{Amariti:2019vfv} and in the case of spherical slicing the only difference is that instead of the Poincar\'e patch of AdS the bulk geometry asymptotes to the global AdS space-time. In the UV limit,  the induced geometry will be again de Sitter with the corresponding Hubble parameter to be governed by the UV limits of the coupling potentials induced on the brane, and thus we will not repeat the analysis here. We are mainly interested in the IR analysis, which is revealed to differ substantially from that of \cite{Amariti:2019vfv}, since in our case the IR geometry is that of a black hole horizon. In particular we find that the induced geometry on the brane in the IR corresponds to the Poincar\'e patch of dS for all the possible slicings. Moreover the Hubble parameter and the scale factor of the induced metric depend on the horizon values of the induced potentials on the brane as well as on the first derivative of the blackening factor which is related to the temperature of the bulk black hole. The most interesting result of our analysis is a relation between the induced cosmology horizon entropy and temperature with the same quantities in the bulk black hole side. We believe that this is a point that certainly deserves more study, since it is quite hard to assign appropriate microstates to the $dS$ entropy, while our analysis indicates that such an understanding could arise from realising $dS$ in a purely QFT theoretical setup (once we understand in more detail the microscopic theory that is dual to such braneworld geometries\footnote{Such a setup can be based on the ideas presented in \cite{smgrav,bbkn}.}).

The structure of the paper is the following: In section \ref{self} we provide the general formulae of the self tuning theory and in section \ref{BH} we describe the matching conditions on the brane in the case of flat, spherical and hyperbolic slicing. In section \ref{probe} we solve these equations analytically in the probe limit where the brane does not backreact on the bulk geometry. We conclude with section \ref{asym} where we present the brane-world cosmology in the IR limit which in our case is captured by the near horizon region of the higher dimensional bulk black hole. This cosmology is that of de-Sitter space (dS), and this allows a natural relation between the entropy and temperature $S_{dS}$ and $T_{dS}$, with those of the higher dimensional bulk black hole $S_{BH}, \, T_{BH}$.
Our paper is complemented with various appendices, where supplementary material as well as more detailed calculations are presented.

\section{The Self-tuning theory}\label{self}

In this section we review the self-tuning mechanism of \cite{selftuning} and \cite{Amariti:2019vfv}.  Our bulk theory is an Einstein - Dilaton theory in $d+1$-dimensions. The bulk space-time coordinates are $x^a\equiv (r, x^\mu)$. Moreover, we embed in the $d+1$ bulk a $d$-dimensional brane that is parametrized by $x^\mu$ . For such a system the most general 2-derivative action is
\be \label{A1}
S = S_{bulk} + S_{brane }
\ee
with,
\be\label{A2}
S_{bulk} = M^{d-1} \int d^{d+1}x \sqrt{-g} \left[R - {1\over 2}g^{ab}\de_a\f\de_b \f - V(\f)\right] + S_{GH},
\ee
\be\label{A3}
S_{brane} = M^{d-1} \int d^d x \sqrt{-\g}\left[- W_B(\f) - {1\over 2} Z_B(\f) \gamma^{\mu\nu}\de_\mu\f\de_\nu \f + U_B(\f) R^{(\gamma)} \right]+\cdots,
\ee
$S_{GH}$ is the Gibbons-Hawking term at the space-time boundary (e.g. the UV boundary if the bulk is asymptotically $AdS$), $M^{d-1} \equiv (16 \pi G_5)^{-1}$ is the bulk Planck scale,  $g_{ab}$ is the bulk metric, $R$ is its associated Ricci scalar, $V(\f)$ is some bulk scalar potential, $\gamma_{\mu\nu}$ is the induced metric on the brane, $R^{(\g)}$ is the intrinsic curvature of the brane.

The ellipsis in \eqref{A3} represent higher derivative terms of the gravitational sector fields ($\f,\gamma_{\m\n}$) as well as the action of the brane-localized fields (such as the ``Standard Model'' (SM)).
 $W_B(\f), Z_B(\f)$ and $U_B(\f)$ are scalar potentials that are localized on the brane since they are generated by the quantum corrections of the brane-localized fields \cite{smgrav}.  For example, $W_B(\f)$ contains  the brane vacuum energy, which takes contributions from the brane matter fields.  All of $W_B(\f), Z_B(\f)$ and $U_B(\f)$ are cutoff dependent and they scale as, $W_B(\f)\sim \Lambda^4$, $Z_B(\f)\sim U_B(\f)\sim \Lambda^2$ where $\Lambda$ is the UV cutoff of the brane physics as described here. Its origin was motivated and described in \cite{selftuning}. Before proceeding with an analysis of the EOM's stemming from eqns. \eqref{A2}, \eqref{A3}, we should also mention that one could extend the braneworld description of eqn. \eqref{A3}, to be that of a quasi-localised brane having a small finite extend in the radial direction. This is natural from a holographic RG point of view, in case that the number of degrees of freedom dual to the braneworld are comparable to those of the bulk gravity. Some similar arguments were also presented in \cite{Fichet:2019owx}. Nevertheless, in the present work we will study the leading contribution ($N \rightarrow \infty, \, M_P \rightarrow \infty$), where the braneworld is exactly localised in the radial direction.

\subsection{Field equations and matching conditions}

The bulk equations  of motion are
\bea
&& R_{ab} -{1\over 2} g_{ab} R = {1\over 2}\de_a\f\de_b \f -  {1\over 2}g_{ab}\left( {1\over 2}g^{cd}\de_c\f\de_d \f + V(\f) \right),  \label{FE1}\\
&& \de_a \left(\sqrt{-g} g^{ab}\de_b \f \right)- \sqrt{-g} {\de V \over \de \f} =0 \label{FE2}
\eea
notice that they depend only on $V(\f)$. The $d$-dimensional brane separates the bulk into two regions. The UV region is the one that extends from the point $r_{0}$ that the d dimensional brane is located and ends on the UV boundary of the $d+1$ bulk spacetime (where the volume form becomes infinite). The IR region is the one that extends from the point $r_{0}$ that the d dimensional brane is located and ends at the interior of the bulk space time where the volume form becomes zero. For the case of bulk black hole geometries that we study in this paper the IR region includes the black hole horizon and ends at the black hole singularity.

We symbolise by  $g^{UV}_{ab}, g^{IR}_{ab}$ and $\f^{UV}, \f^{IR}$  the solutions for the metric and scalar field in the UV and IR regions of the brane respectively. The $\Big[ X\Big]^{IR}_{UV}$ symbolise the ``jump'' of a quantity $X$ across the brane. Then, we can express Israel's junction conditions as
\begin{enumerate}
\item A continuity equation for the metric and scalar field:
\be\label{FE3}
\Big[g_{ab}\Big]^{UV}_{IR} = 0,   \qquad \Big[\f\Big]^{IR}_{UV} =0
\ee
\item The extrinsic curvature and the normal derivative of $\f$ should satisfy the following discontinuity conditions:
\be\label{FE4}
\Big[K_{\mu\nu} - \gamma_{\mu\nu} K \Big]^{IR}_{UV} =   {1\over \sqrt{-\gamma}}{\delta S_{brane} \over \delta \gamma^{\mu\nu}}  ,  \qquad \Big[n^a\de_a \f\Big]^{IR}_{UV} =- {1\over \sqrt{-\gamma}}{\delta S_{brane} \over \delta \f} ,
\ee
here $\gamma$ is the induced metric on the brane, $K_{\mu\nu}$ is the extrinsic curvature of the brane, $K = \gamma^{\mu\nu}K_{\mu\nu}$  its trace, and $n^a$ a unit normal vector to the brane,  pointing  into the $IR$.
\end{enumerate}
The discontinuity equations \eqref{FE4} for an action of the form \eqref{A3} are given explicitly by
\bea
&&
\Big[K_{\mu\nu} - \gamma_{\mu\nu} K \Big]^{IR}_{UV} = \Bigg[\ha W_B(\f) \gamma_{\mu\nu} + U_B(\f) G^{(\gamma)}_{\mu\nu} - {1\over 2}Z_B(\f) \left(\de_\mu \f \de_\nu \f - \ha \g_{\mu\nu} (\de \f)^2 \right)  \nonumber \\
&& \qquad + \left(\gamma_{\mu\nu}\gamma^{\rho\sigma}\nabla^{(\g)}_\rho \nabla^{(\g)}_\sigma - \nabla^{(\g)}_\mu\nabla^{(\g)}_\nu\right) U_B(\f) \Bigg]_{\f_0(x)}, \label{FE5}\\
&& \Big[n^a\de_a \f\Big]^{IR}_{UV}  = \left[{d W_B \over d \f} - {d U_B\over d \f} R^{(\g)} + \ha {d Z_B \over d \phi}(\de \f)^2 - {1\over \sqrt{\g}}\de_\mu \left( Z_B \sqrt{\g} \g^{\mu\nu}\de_\nu \f \right) \right]_{\f_0(x)},  \label{FE6}
\eea
where $\f_0(x^\mu)$ is the scalar field on the brane.

\section{Embeddings of Black Holes}\label{BH}
In this section we present the specific form that the relations \eqref{FE5}, \eqref{FE6} take in the presence of bulk black holes with non-trivial scalar field profile. In the following subsections we examine in detail the case of flat \ref{EFH} and spherical \ref{ESH} slicing and in Appendix \ref{hyp} we derive the equations in the case of hyperbolic slicing.

\subsection{Embedding of a Black hole with  flat slicing}\label{EFH}
In this subsection we examine what are the conditions \eqref{FE5}, \eqref{FE6} in the case that the bulk geometry is a black hole with flat slicing with a metric of the form

\be
ds^2={dr^2\over f(r)}+e^{2A(r)}\left[-f(r)dt^2+dx_{i}dx^{i}\right],
\label{e1}\ee
 $f(r)$ is the blackening factor  and $e^{2A(r)}$ is the scale factor of the metric. In order to compute the extrinsic curvature $K_{\mu\nu}$ and its trace $K$ it is useful to bring the metric \eqref{e1} into an ADM form where we use as ``time'' the radial bulk direction and decompose the metric as
\be\label{HJ1}
d s^2 = g_{M N} d x^M dx^N = {\gamma}_{\m\n} d  x^\m d  x^\n + 2 N_\m d  x^\m d r + (N^2 + N_\m N^\m) d r^2 \, ,
\ee
with $N, \, N_\m$ the lapse and shift functions and $\gamma_{\m \n}$ the induced metric on the hypersurfaces $\Sigma_r$.

Using the unit normal vector on the surface $n^M =  \left(1/N,- N^\mu /N \right)$\footnote{$\sigma = n^M n_M =  +1 $, or $-1$ if the surface is timelike/spacelike (which is a possibility if the bulk theory is Lorentzian), in our case $\sigma = n^M n_M =  +1$.} we define the extrinsic curvature and its trace as
\be\label{HJ2}
K_{\m \n} = \half (\mathcal{L}_n g)_{\m \n} = \frac{1}{2 N} \left( \dot \gamma_{\m \n} - \nabla_\m N_\n - \nabla_\n N_\m  \right) \, , \quad K = K_{\m \n} \gamma^{\m \n}
\ee

For a metric of the form \eqref{e1}, we have that
\be \label{a2}
N=\frac{1}{\sqrt{f(r)}},\quad N_{\mu}=0,\quad \gamma_{ii}=e^{2A},\quad\gamma_{tt}=-e^{2A}f(r)\,.
\ee

The extrinsic curvature components using \eqref{HJ2} and \eqref{a2} are

\be \label{Ktt}
K_{tt}=-\frac{\sqrt{f(r)}}{2}\left(2\dot{A} f(r)+\dot{f(r)}\right)e^{2A}
\ee

\be \label{Kii}
K_{ij}=\sqrt{f(r)}\dot{A} e^{2A}\delta_{ij}
\ee

and the trace of the extrinsic curvature is

\be \label{KT}
K=\sqrt{f(r)}\left[d\, \dot{A}+\frac{1}{2}\frac{\dot{f(r)}}{f(r)}\right].
\ee

By substituting \eqref{Ktt}, \eqref{Kii}, \eqref{KT} at  \eqref{FE5} we find

\be\label{c1}
\left[K_{tt}-\gamma_{tt}K\right]_{UV}^{IR}=(d-1)e^{2A}f(r)^{3/2} \left[ \dot{A}\right] _{UV}^{IR}=-\frac{1}{2} W_{B}e^{2A}f(r)
\ee
or
\be \label{cond1}
W_{B}=-2(d-1)\sqrt{f(r)}\left[ \dot{A}\right] _{UV}^{IR}
\ee
and
\be\label{c1K}
\left[K_{ii}-\gamma_{ii}K\right]_{UV}^{IR}=-e^{2A}\sqrt{f(r)}\left[(d-1)\dot{A}+\frac{1}{2}\frac{\dot{f(r)}}{f(r)}\right]_{UV}^{IR}=\frac{1}{2} W_{B}e^{2A}
\ee
or
\be\label{cond2}
W_{B}=\left[-2(d-1)\sqrt{f(r)}\dot{A}-\frac{\dot{f(r)}}{\sqrt{f(r)}}\right]_{UV}^{IR}\,.
\ee
Moreover \eqref{FE6} becomes
\be \label{cond3}
n^{\alpha} \partial_{a} \varphi\vert_{UV}^{IR}=\sqrt{f(r)} \left[ \partial_{r}\varphi\right]_{UV}^{IR}=\frac{dW_{B}}{d\varphi}\,.
\ee

For stabilization of the brane to occur at some point $r_{0}$ in the bulk five dimensional spacetime, the conditions \eqref{cond1}, \eqref{cond2} and \eqref{cond3} should be satisfied simultaneously. That is the case only if

\be
\left[\frac{\dot{f(r)}}{\sqrt{f(r)}}\right]_{UV}^{IR}=0
\ee
which means that the derivative of the blackening factor should be continuous before and after the location of the brane $r_{0}$. Using the superpotential formalism described in Appendix \ref{ASF1} we can rewrite the equations \eqref{cond1} - \eqref{cond3} as

\be \label{cond1b}
W_{B}=\sqrt{f(\f(r))}\left(W_{IR}- W_{UV}\right)
\ee

\be \label{cond2b}
W_{B}=\sqrt{f(\f(r))}\left(W_{IR}- W_{UV}\right)-\left[\frac{f^{\prime} W^{\prime}}{\sqrt{f(\f(r))}}\right]_{UV}^{IR}
\ee

\be \label{cond3b}
W_{B}^{\prime}=\sqrt{f(\f(r))}\left(W_{IR}^{\prime}- W_{UV}^{\prime}\right)
\ee
where $\prime$ symbolise derivatives with respect to $\f$. Both $W_{IR},\, W_{UV}$ are solutions to the superpotential equation

\be
\left(\frac{d }{4 (d-1)}W^2(\f)-\frac{(W'(\f))^2}{2}\right)f(\f)-{W'\over 2}f'(\f) W(\f)+V(\f)=0.
\label{eW30}\ee

\paragraph{Holographic self-tuning}

Following \cite{selftuning}, we use the rules of holography to fix the integration constant $C_{IR}$,  of $W_{IR}$  by demanding regularity in the $IR$\footnote{Usually, there is just one such solution to \eqref{eW30} or a discrete set, thus $W_{IR}$ is fixed by the regularity condition.}.  Then the matching conditions \eqref{cond1b}-\eqref{cond3b} fix the integration constant $C_{UV}$ for the UV superpotential $W_{UV}$, the position of the brane $\f_{0}$ in the field space and how the blackening factor $f(\f(r))$ behaves around $\f_{0}$.

The integration constant $C_{UV}$ through the holographic dictionary is related to the VeV of the operator dual to $\f$, thus the effect of the insertion of the brane is to change the VeV comparatively to the case without brane \footnote{Of course we can still freely specify the UV sources for our fields through the UV boundary conditions imposed on the bulk equations of motion.}.

In order to have self-tuning of the CC and not a fine tuning, one should be able to find the 4d Minkowski space geometry on the brane for generic values of the parameters. In the case at hand, the parameters of the model are the bulk and brane potentials, which contain the 4-$d$ vacuum energy\footnote{It is included in the $\f$-independent part of $W_B(\f)$.}. 

In more detail the process is the following, for arbitrary values of the potentials $V$ and $W_B$ in the bulk and on the brane, the UV side of the geometry adjusts itself dynamically, for given $C_{UV}$ and $\f_0$, so that the induced geometry on the brane can be that of 4- $d$ Minkowski. We observe that for arbitrary initial conditions for $W_{UV}$ at $\f_0$, the space-time at $\f_{0}$ connects to the same UV AdS region. Consequently,  we conclude that any value of  $W_{UV}$ gives rise to a regular geometry that satisfies the same boundary conditions. 

From the boundary field theory point of view, these geometries are distinct only due to the different VeV of the operator dual to $\f$. This VeV is related to the integration constant $C_{UV}$ that fixes $W_{UV}$. We conclude that  the UV geometry self-adjusts such that it can be pasted to the regular IR solution on the brane at $\f_{0}$ for any value of the parameters there.

\subsection{Embedding of a Black hole with spherical slicing}\label{ESH}

In this subsection we study the case of the bulk geometry to be a black hole with spherical slicing using the following ansatz for the metric
\be
ds^2={dr^2\over f(r)}+e^{2A(r)}\left[-f(r)dt^2+R^2 d\Omega_{d-1}^{2}\right].
\label{m1}\ee
where $f(r)$ is the blackening factor, $e^{2A(r)}$ is the scale factor of the metric, $d\Omega_{d-1}$ is the metric of the unit transverse sphere and $R$ is the radius of the transverse sphere.
This ansatz is appropriate when the dual QFT to the bulk theory is defined on an $R\times S^{d-1}$ geometry.

In the case of five bulk dimensions the explicit metric for the unit sphere is
\be
d\Omega_{3}^{2}=d\psi^2+\sin^2\psi\left(d\theta^{2}+\sin^{2}\theta\,d\phi^2\right).\ee
In order to compute the extrinsic curvature and its trace it is convenient to use the ADM form of the metric \eqref{HJ1} with the following identifications

\be\label{NS}
N=\frac{1}{\sqrt{f(r)}},\quad N_{\mu}=0,\quad \gamma_{tt}=-e^{2A}f(r),\quad \gamma_{ij}=e^{2A}R^2\chi_{ij}(\Omega_{d-1}),
\ee
where $\chi_{ij}$ are the metric components of the $d-1$ dimensional unit sphere\footnote{For the case of five bulk dimensions the transverse sphere is a three sphere and the explicit components of the induced metric are $\gamma_{\psi\psi}=e^{2A} R^2$, $\gamma_{\theta\theta}=e^{2A} R^2 \sin^{2}\psi$ and $\gamma_{\theta\theta}=e^{2A} R^2 \sin^{2}\psi\,\sin^{2}\theta$.}.

Substituting \eqref{NS} in \eqref{HJ2} the components of the extrinsic curvature are

\be\label{KttS}
K_{tt}=-\frac{\sqrt{f(r)}}{2}\left(2\dot{A} f(r)+\dot{f(r)}\right)e^{2A}
\ee

\be \label{KiiS}
K_{ij}=\sqrt{f(r)}\dot{A} e^{2A}R^2\chi_{ij}(\Omega_{d-1})
\ee
and the trace is

\be \label{KTS}
K=\sqrt{f(r)}\left[d\, \dot{A}+\frac{1}{2}\frac{\dot{f(r)}}{f(r)}\right].
\ee
The Ricci scalar induced on the brane is
\be\label{Ricci}
R^{\gamma}=\frac{(d-2)(d-1) e^{-2 A}}{R^2}.
\ee

The Einstein tensor induced on the brane can be written in the compact form\footnote{In the case of a four dimensional brane the explicit form of the Einstein tensor is \be
G^{\gamma}_{\mu\nu}=\begin{pmatrix}
\frac{3 f(r)}{R^2} & 0 & 0 & 0\\
0 & -1 & 0 & 0 \\
0 & 0 & -\sin^2 \psi  & 0\\
0 & 0 & 0 & -\sin^2 \psi \sin^2 \theta
\end{pmatrix}
\ee}

\be \label{Einstein}
G^{\gamma}_{\mu\nu}=\text{diagonal} \left\lbrace \frac{(d-1)f(r)}{R^2}, -\chi_{ii}  \right\rbrace\,.
\ee

By substituting \eqref{KttS}, \eqref{KiiS}, \eqref{KTS} \eqref{Ricci} and \eqref{Einstein} in \eqref{FE5} we find the following two relations

\be\label{d1}
\left[K_{tt}-\gamma_{tt}K\right]_{UV}^{IR}=(d-1)e^{2A}f(r)^{3/2}\left[ \dot{A}\right]_{UV}^{IR}=-\frac{1}{2} W_{B}e^{2A}f(r) +U_{B}(\varphi)(d-1) \frac{f(r)}{R^2}
\ee

\be\label{d2}
\left[K_{ij}-\gamma_{ij}K\right]_{UV}^{IR}=-e^{2A}R^2\chi_{ij}(\Omega_{d-1})\sqrt{f(r)}\left[(d-1)\, \dot{A}+\frac{1}{2}\frac{\dot{f(r)}}{f(r)}\right]_{UV}^{IR}
\ee
$$=\frac{1}{2} W_{B}e^{2A} R^2 \chi_{ij}(\Omega_{d-1})-U_{B}(\varphi)\chi_{ij}(\Omega_{d-1}).$$

Moreover, \eqref{FE6} becomes
\be \label{d3}
n^{\alpha}\partial_{a}\varphi\vert_{UV}^{IR}=\sqrt{f(r)} \left[ \partial_{r}\varphi\right]_{UV}^{IR}=\frac{dW_{B}}{d\varphi}-\frac{e^{-2A}(d-2)(d-1)}{R^2}\frac{d U_{B}}{d\varphi}
\ee

For stabilization of the brane to occur at some point $r_{0}$ in the bulk five dimensional spacetime, the conditions \eqref{d1}, \eqref{d2} and \eqref{d3} should be satisfied simultaneously. That is the case only if
\be \label{con1}
\frac{U_{B}(d-2)}{R^2}=-\frac{e^{2A}}{2}\left[ \frac{\dot{f(r)}}{\sqrt{f(r)}}\right]_{UV}^{IR}\,.
\ee

Additionally from \eqref{A3}, we observe that in order to have a positive Planck scale on the brane, $U_{B}$ should be positive\footnote{A special case is $U_{B}=0$ when $\dot{f(r)}$ is continuous. In such a case the conclusions of subsection \ref{EFH} hold.} and thus
\be\label{con2}
\dot{f(r)}_{IR}-\dot{f(r)}_{UV}\vert_{r=r_{0}}<0\, ,
\ee
which means that the derivative of the blackening factor should be discontinuous at the point $r=r_{0}$ where the brane is located and decreasing from the UV ($r_0+\epsilon,\, \epsilon>0$) to the IR  ($r_0-\epsilon,\, \epsilon>0$).  Using the superpotential formalism presented in Appendix  \ref{BSF1} we can rewrite the equations \eqref{d1} - \eqref{d3} as

\be
\sqrt{f(\f(r))}\left(W_{IR}- W_{UV}\right)=W_{B}-2e^{-2A}U_{B}(\varphi)(d-1) \frac{1}{R^2}
\ee

\be
\sqrt{f(\f(r))}\left(W_{IR}- W_{UV}\right)-\left[\frac{f^{\prime} W^{\prime}}{\sqrt{f(\f(r))}}\right]_{UV}^{IR}=W_{B} -2 e^{-2A}U_{B}(\varphi)\frac{1}{R^2}
\ee

\be
W_{B}^{\prime}-\frac{e^{-2A}(d-2)(d-1)}{R^2}U_{B}^{\prime}= \sqrt{f(\f(r))}\left(W_{IR}^{\prime}- W_{UV}^{\prime}\right)
\ee
where the $\prime$ are derivatives with respect to $\f$ and
\be
A(\f)=-\frac{1}{2(d-1)} \int_{\f^*}^\f \frac{W(\f')}{W'(\f')} d \f'.
\label{w54}\ee

Moreover, both $W_{IR},\, W_{UV}$ are solutions to the superpotential equation
\be
\bigg( \frac{d}{4(d-1)} W^2(\f)-\frac{(W'(\f))^2}{2}\bigg) f(\f)-\frac{1}{2} W'(\f)W(\f)f'(\f)+V(\f)-(d-1)(d-2)T(\f)=0.
\label{w55}\ee

The discussion presented in the subsection \ref{EFH} about the holographic self-tuning, follows similarly here. Additionally, in this case we have one more parameter $U_{B}$ on the brane that is related through \eqref{con1} to the discontinuity of the blackening factor at the point where the brane is inserted.

\subsection{Embedding of a Black Hole with hyperbolic slicing}
The case of a bulk black hole with hyperbolic slicing is presented in detail in the appendix \ref{hyp}. We briefly state here, that one follows the exact same steps as for the case of spherical slicing, the only difference in the present case is that the curvature \eqref{Ricci} has an overall minus sign. Due to this sign flip in the curvature, the equivalent equation to \eqref{d3} in the hyperbolic case is
\be \label{dh3}
n^{\alpha}\partial_{a}\varphi\vert_{UV}^{IR}=\sqrt{f(r)} \left[ \partial_{r}\varphi\right]_{UV}^{IR}=\frac{dW_{B}}{d\varphi}+\frac{e^{-2A}(d-2)(d-1)}{R^2}\frac{d U_{B}}{d\varphi}\,.
\ee
The equations \eqref{d1}, \eqref{d2} and \eqref{con1}, \eqref{con2},  as well as the holographic self-tuning procedure remain the same.

To conclude this section we comment on the possibility of addressing the full backreacting problem. The bulk equations of motion \eqref{FE1} and \eqref{FE2} supplemented by the junction conditions \eqref{FE5} and \eqref{FE6} can be solved analytically only in the static case as was done in \cite{selftuning}. In the present note we are interested in the induced cosmology on the brane thus the metric on the brane should be time dependent.  Unfortunately, in such a case one cannnot solve the equations analytically, only numerically. Since we want to have analytic control on our results, in the following we constrain our analysis in the probe limit where the brane does not backreact on the bulk geometry.

\section{The probe brane limit}\label{probe}

In this section we study the brane dynamics in the probe limit. In this limit one has to solve the bulk equations of motion \eqref{FE1} and \eqref{FE2} without taking into account the backreaction of the brane in the bulk geometry\footnote{As we saw in section \ref{self} for a self-tuning mechanism to operate, the brane should backreact on the bulk geometry. In order to take the probe limit we need a regime of parameters such that the induced action of the brane is much smaller than the bulk one, hence the brane cosmological term cannot be big. Despite this shortcoming our analysis can still offer a qualitative understanding of self-tuning cosmologies as was discussed in \cite{Amariti:2019vfv}.}. This translates into accepting that the bulk geometry is smooth across the brane and we simply have to set the \eqref{FE4} equal to zero. Consequently, the bulk geometries in our examples are given by \eqref{e1} and \eqref{m1} respectively.

The induced action on the brane  before gauge fixing, in this regime is
\be
S_b= M^3 \int d^4\xi\sqrt{-\hat g}\left(-W_B(\varphi)+U_B(\varphi)\hat R-{Z_B\over 2}(\pa\varphi)^2+\cdots\right)\;,
\label{c2}\ee
where $\xi^{\m}$ are world-volume coordinates and the hat indicates induced quantities.

\subsection{Flat Slicing}\label{fS1}

In this subsection we study the probe brane limit in the case where the bulk black hole geometry has a flat slicing and is described by the metric \eqref{e1}. Moreover, we work in the static gauge, $\xi^{\m}=x^{\mu}$, and hence the only dynamical variable is $r(x^{\m})$ that we allow it only to depend on time. Thus the brane in the probe limit has only one dynamical degree of freedom namely $r(t)$. The induced metric on the brane is given by
\be
d\hat s^2\equiv \hat g_{\m\n}d\xi^{\m}d\xi_{\n}=
-\left(e^{2A}  f(r)-\frac{\dot r^2}{f(r)}\right) dt^2+e^{2A}dx^idx^i
\label{c3}\ee
where in the rest of this section a dot denotes: $\dot{~}\equiv\frac{d}{dt}$.

The induced Ricci scalar on the brane times the square root of the determinant of the induced metric is found to be

\be \label{R1}
\sqrt{-\hat{g}} \hat{R} =\frac{6 e^{3A}\sqrt{f}\left(\ddot{A}+\dot{A}^2\right)}{\left(e^{2A}f^2-\dot{r}^2\right)^{1\over2}}+\frac{6e^{3A} \sqrt{f}\left(-\dot{A}^2 \dot{r}^2 + \dot{A}\dot{r}\ddot{r}\right)}{\left(e^{2A}f^2-\dot{r}^2\right)^{3\over2}}-\frac{3e^{3A}\dot{A}\dot{f}\left(e^{2A}f^2 +\dot{r}^2\right)}{\sqrt{f}\left(e^{2A}f^2-\dot{r}^2\right)^{3\over2}}\,.
\ee

We then foliate the induced metric with spacelike surfaces of constant time. The extrinsic curvature and its trace of these surfaces can be computed to yield
\be\label{K}
K_{ij}=\frac{\delta_{ij}\sqrt{f}\dot{e^{2A}}}{2\left(e^{2A}f^2-\dot{r}^2\right)^{1\over2}} \, , \quad K=\frac{3\dot{A}\sqrt{f}}{\left(e^{2A}f^2-\dot{r}^2\right)^{1\over2}} \,.
\ee
Substituting then \eqref{K} into \eqref{R1} we have
\be
\sqrt{-\hat{g}} \hat{R} = \frac{-6 e^{3A}\sqrt{f}\dot{A}^2}{\left(e^{2A}f^2-\dot{r}^2\right)^{1\over2}}+2 \partial_{t}\left(\sqrt{\hat{\gamma}}K\right)
\ee
where $\hat{\gamma}_{ij}=\delta_{ij} e^{2A},\,{i,j=1,...,3}$.

One can then express the action of the brane \eqref{c2} as
\be
S_b=- M^3 V_{3} \int dt\left[\sqrt{\left(e^{2A}f^2-\dot{r}^2\right)}\frac{e^{3A}}{\sqrt{f}}\left(W_B(\varphi)+{Z_B\over 2}g^{tt}\dot{\varphi}^2\right)\right]
\label{c22}\ee
$$
- M^3 V_{3} \int dt\,e^{3A}\,\left(\dot{U}_{B}(\varphi)+U_B(\varphi)\dot{A}\right)\,\frac{6 \sqrt{f}\dot{A}}{\left(e^{2A}f^2-\dot{r}^2\right)^{1\over2}}
$$
where we cancelled the total derivative term with the Gibbons-Hawking term at the boundaries of the time integral. The action is then reduced to the one of a one dimensional Lagrangian mechanics problem.

Using the definition for the superpotential \eqref{e26}
\be \label{e126}
\dot{r} W(\phi) \equiv -6 \dot{A}\,,\quad \dot\phi=W'\dot{r}\,,\quad \dot{U}_{B}(\varphi)=U_{B}^{\prime} W^{\prime}\dot{r}
\ee
we can write the action \eqref{c22} as
%\be
%S_b[r(t)]=-M^3 V_3\int\left[ dt~e^{4A}\sqrt{f}\sqrt{ 1-\frac{e^{-2A}\dot r^2}{f^2}}W_B-\frac{\dot{r}^2 e^{2A}}{\sqrt{f}\sqrt{ 1-\frac{e^{-2A}\dot r^2}{f^2}}}F\right]
%\equiv \int dt~L_b
%\label{c122}\ee
\be
S_b[r(t)]=-M^3 V_3\int dt~e^{4A}\sqrt{f}\left[ \sqrt{ 1-\frac{e^{-2A}\dot r^2}{f^2}}\left(W_B+ F f\right)-\frac{F f}{\sqrt{ 1-\frac{e^{-2A}\dot r^2}{f^2}}}\right]
\equiv \int dt~L_b \, ,
\label{c1222}\ee
where $V_3$ is the spatial volume and
\be
F(\varphi)=-{U_BW^2\over 6}+W {d W\over d\f} {d U_B \over d\f}+{1\over
  2}Z_B\left({d W \over d\f}\right)^2 ~.
\label{c13}\ee

\subsubsection{The general solution for flat slicing}

The Hamiltonian corresponding to the action \eqref{c1222} is conserved since the action \eqref{c1222} is explicitly time-independent. The momentum conjugate to $r$ is
\be
p_r={\delta L_b\over \delta \dot r}={e^{2A}\dot r\sqrt{f}\over{f^2 \left( 1-\frac{e^{-2A}\dot r^2}{ f^2}\right)^{3\over 2}}}\left[F f + \left( 1-e^{-2A}\dot r^2\over f^2\right) (W_B+F f)\right]
\label{c14}\ee
and the conserved Hamiltonian is
\be
H=\dot r p_r-L_b=\frac{e^{2A}\sqrt{f}}{ ( 1-\frac{e^{-2A}\dot{r}^2}{f^2})^{3\over 2}} \left[e^{2A}W_B+(F f-W_B)\frac{\dot r^2}{ f^2}\right] =E\;,
\label{c15}\ee
where we have reabsorbed the overall $M^3V_3$ into a new one, namely $E$.

Given $E$, we can solve \eqref{c15} for $\dot r$
\be
E( 1-\frac{e^{-2A}\dot{r}^2}{f^2})^{3\over 2}=e^{2A}\sqrt{f}\left[e^{2A}W_B+(F f -W_B)\frac{\dot r^2}{f^2}\right]\,.
\label{c29}\ee
From the solution of \eqref{c29} we obtain $\dot r^2$ as a function of $e^{2A(r)}, \, f(r)$ and $\varphi(r)$. Moreover, by performing the following coordinate transformation in (\ref{c3}) we obtain
\be
\sqrt{e^{2A}f-\frac{\dot r^2}{f}}~dt=d\tau
\quad
\rightarrow
\quad
\frac{d r}{ d t}
=
\frac{\sqrt{f} e^A}{\sqrt{1+\frac{1}{f}\left(\frac{d r}{d \tau}\right)^2}}
\frac{d r}{ d \tau}\;.
\label{cc1}
\ee
Then, we can write the induced metric on the brane as
\be
d\hat s^2\equiv -d\tau^2+e^{2A(\tau)}dx^idx^i
\label{c32}\ee
where $\tau$ is the proper time on the brane. Finally, by introducing the new variable $y$
\begin{equation} \label{y}
y \equiv  \sqrt{1+\frac{1}{f}\left(\frac{d r}{d \tau}\right)^2},\quad \text{with}\quad y\geq 1
\end{equation}
we can express \eqref{c29} in this simpler form
\begin{equation}
E  = e^{4 A} y\sqrt{f}[F f (y^2-1)+W_B] \, .
\label{cc33}
\end{equation}
This is a cubic equation for $y$, that can be solved according to the analysis of Appendix \ref{AppCubic}. Consequently, we have translated the problem of finding the trajectory of the probe brane in the bulk geometry into solving \eqref{cc33} for $y(r)$ for fixed $E$\footnote{$E$ acts as an initial condition for the brane's trajectory. The second initial  condition is just a shift of the initial time point.} taking into account that $y(r)\geq 1$.

\subsection{Spherical Slicing}\label{sS1}

In this subsection we study the probe brane limit in the case where the bulk black hole geometry has a spherical symmetry and is described by the metric \eqref{m1}. Again, we work in the static gauge, $\xi^{\m}=x^{\mu}$, and hence the only dynamical variable is $r(x^{\m})$ that we allow it only to depend on time. Thus the brane in the probe limit has only one dynamical degree of freedom namely $r(t)$. The induced metric on the brane is given by
\be
d\hat s^2\equiv \hat g_{\m\n}d\xi^{\m}d\xi_{\n}=
-\left(e^{2A}  f(r)-\frac{\dot r^2}{f(r)}\right) dt^2+e^{2A}R^2 d\Omega_{d-1}^{2}
\label{cS3}\ee
where $d\Omega_{d-1}$ is the metric of the transverse unit sphere. In the case of five bulk dimensions the explicit metric is
\be
d\Omega_{3}^{2}=d\psi^2+\sin^2\psi\left(d\theta^{2}+\sin^{2}\theta\,d\phi^2\right)\,.
\ee
The induced Ricci scalar on the brane times the square root of the determinant of the induced metric is found to be
\be \label{RS1}
\sqrt{-\hat{g}}\hat{R} =6 R^3 e^{3A}\sin^{2}\psi \sin\theta\left[\frac{\sqrt{f}\left(\ddot{A}+\dot{A}^2\right)}{\left(e^{2A}f^2-\dot{r}^2\right)^{1\over2}}+\frac{ \sqrt{f}\left(-\dot{A}^2 \dot{r}^2 + \dot{A}\dot{r}\ddot{r}\right)}{\left(e^{2A}f^2-\dot{r}^2\right)^{3\over2}}-\frac{\dot{A}\dot{f}\left(e^{2A}f^2 +\dot{r}^2\right)}{2\sqrt{f}\left(e^{2A}f^2-\dot{r}^2\right)^{3\over2}}\right]+
\ee
$$
+6 R^3 e^{3A}\sin^{2}\psi \sin\theta\left[\frac{e^{-2 A}\left(e^{2A}f^2-\dot{r}^2\right)^{1\over2}}{R^2 f^{1/2}}\right]
$$

We then foliate the induced metric with surfaces of constant time. The extrinsic curvature and its trace are
\be\label{KS}
K_{ij}=\frac{R^2\chi_{ij}\sqrt{f}\dot{e^{2A}}}{2\left(e^{2A}f^2-\dot{r}^2\right)^{1\over2}} \, , \quad K=\frac{3\dot{A}\sqrt{f}}{\left(e^{2A}f^2-\dot{r}^2\right)^{1\over2}}\,.
\ee
Substituting \eqref{KS} in \eqref{RS1} we have
\be
\sqrt{-\hat{g}} \hat{R} =6 e^{3A} R^3 \sin^{2}\psi \sin\theta\left[\frac{e^{-2 A}\left(e^{2A}f^2-\dot{r}^2\right)^{1\over2}}{R^2 f^{1/2}}-\frac{\sqrt{f}\dot{A}^2}{\left(e^{2A}f^2-\dot{r}^2\right)^{1\over2}}\right]+2 \partial_{t}\left(\sqrt{\hat{\gamma}}K\right)
\ee
where $\hat{\gamma}_{ij}=\chi_{ij}R^2 e^{2A},\,{i,j=1,...,3}$ and the action \eqref{c2} can be written as
\be
S_b= M^3 V_{3} \int dt\left[\sqrt{\left(e^{2A}f^2-\dot{r}^2\right)}\frac{e^{3A}}{\sqrt{f}}\left(-W_B(\varphi)+ 6\frac{U_B e^{-2A}}{R^2}\right)\right]+
\label{cS22}\ee
$$
+M^3 V_{3} \int dt\,e^{3A}\,\left[\frac{Z_B}{2}\dot{\f}^2-6 \dot{A}\left(\dot{U}_{B}(\varphi)+U_B(\varphi)\dot{A}\right)\right]\,\frac{ \sqrt{f}}{\left(e^{2A}f^2-\dot{r}^2\right)^{1\over2}}
$$
where we cancelled the total derivative term with the Gibbons-Hawking term of the action.

Using the definition for the superpotential \eqref{e26} we have
\be \label{e226}
\dot{r} W(\phi) \equiv -6 \dot{A}\,,\quad \dot\phi=W'\dot{r}\,,\quad \dot{U_{B}(\varphi)}=U_{B}^{\prime} W^{\prime}\dot{r}
\ee
and the action \eqref{cS22} can be written in a more compact form as
\be
S_b[r(t)]=-M^3 V_3\int dt~e^{4A}\sqrt{f}\left[\sqrt{ 1-\frac{e^{-2A}\dot r^2}{f^2}}\left(W_B+Ff-6 \frac{U_{B}e^{-2A}}{R^2}\right)-\frac{Ff}{\sqrt{ 1-\frac{e^{-2A}\dot r^2}{f^2}}}\right]
\label{cS1222}\ee
where $V_3$ is the spatial volume and $F$ is the same as in the flat slicing case, namely
\be
F(\varphi)=-{U_BW^2\over 6}+W {d W\over d\f} {d U_B \over d\f}+{1\over
  2}Z_B\left({d W \over d\f}\right)^2 ~.
\label{c13}\ee

\subsubsection{The general solution for spherical slicing}

We will again use the fact that the Hamiltonian to the action \eqref{cS1222} is conserved since the action \eqref{cS1222} is explicitly time-independent. The momentum conjugate to $r$ is

\be
p_r={\delta L_b\over \delta \dot r}={e^{2A}\dot r\sqrt{f}\over{f^2 \left( 1-\frac{e^{-2A}\dot r^2}{ f^2}\right)^{3\over 2}}}\left[F f + \left( 1-e^{-2A}\dot r^2\over f^2\right) (W_B+F f-6\frac{U_{B}e^{2A}}{R^2})\right]
\label{cS14}\ee

and the conserved Hamiltonian is

\be
H=\dot r p_r-L_b=\frac{e^{2A}\sqrt{f}}{ ( 1-\frac{e^{-2A}\dot{r}^2}{f^2})^{3\over 2}} \left[e^{2A}(W_B-6\frac{U_{B}e^{-2A}}{R^2})+(F f-W_B+6\frac{U_{B}e^{-2A}}{R^2})\frac{\dot r^2}{ f^2}\right] =E\;,
\label{cS15}\ee
where we have reabsorb the overall $M^3V_3$ into a new one, namely $E$.

Given $E$, we can solve \eqref{cS15} for $\dot r$
\be
E( 1-\frac{e^{-2A}\dot{r}^2}{f^2})^{3\over 2}=e^{2A}\sqrt{f} \left[e^{2A}(W_B-6\frac{U_{B}e^{-2A}}{R^2})+(F f-W_B+6\frac{U_{B}e^{-2A}}{R^2})\frac{\dot r^2}{ f^2}\right]
\label{cS29}\ee

From the solution of \eqref{cS29} we obtain $\dot r^2$ as a function of $e^{2A(r)}, \, f(r)$ and $\varphi(r)$. Moreover, by performing the following coordinate transformation in (\ref{cS3}) we obtain
\be
\sqrt{e^{2A}f-\frac{\dot r^2}{f}}~dt=d\tau
\quad
\rightarrow
\quad
\frac{d r}{ d t}
=
\frac{\sqrt{f} e^A}{\sqrt{1+\frac{1}{f}\left(\frac{d r}{d \tau}\right)^2}}
\frac{d r}{ d \tau}\;.
\label{cc1}
\ee
 Then we can write the induced metric on the brane as
 \be
d\hat s^2\equiv -d\tau^2+e^{2A(\tau)}R^2d\Omega_{3}
\label{cS32}\ee
where $\tau$ is the proper time on the brane. Finally, by introducing the new variable $y$

\begin{equation} \label{yS}
y \equiv  \sqrt{1+\frac{1}{f}\left(\frac{d r}{d \tau}\right)^2},\quad \text{with}\quad y\geq 1
\end{equation}

we can write

\be
\dot{r}^2 =\frac{y^2 -1}{y^2}f^2 e^{2A}
\ee
and the equation (\ref{cS29}) simplifies to
\begin{equation}
E  = e^{4 A} y\sqrt{f}[F f (y^2-1)+W_B-6\frac{U_{B}e^{-2A}}{R^2}]
\label{ccS33}
\end{equation}
\eqref{ccS33} is a cubic equation for $y$, and can be solved as shown in appendix \ref{AppCubic}. Consequently, the problem of finding the trajectory of the probe brane in the bulk geometry translates into solving \eqref{ccS33} for $y(r)$ for fixed $E$
and imposing  $y(r)\geq 1$.

\subsection{Hyperbolic Slicing}\label{sH1}

For the bulk black hole with hyperbolic slicing the procedure is exactly the same as in the case of the spherical slicing described in the subsection \ref{sS1}. The induced metric on the brane is now
\be
d\hat s^2\equiv \hat g_{\m\n}d\xi^{\m}d\xi_{\n}=
-\left(e^{2A}  f(r)-\frac{\dot r^2}{f(r)}\right) dt^2+e^{2A}R^2 d\H_{d-1}^{2}
\label{cS3h}\ee
where $d\H_{d-1}$ is the metric of the transverse hyperbolic space. In the case of five bulk dimensions the explicit metric is
\be
d\H_{3}^{2}=d\psi^2+\sinh^2\psi\left(d\theta^{2}+\sinh^{2}\theta\,d\phi^2\right)\,.
\ee

The induced action on the brane is

\be
S_{bH}[r(t)]=-M^3 V_3\int dt~e^{4A}\sqrt{f}\left[\sqrt{ 1-\frac{e^{-2A}\dot r^2}{f^2}}\left(W_B+Ff+6 \frac{U_{B}e^{-2A}}{R^2}\right)-\frac{Ff}{\sqrt{ 1-\frac{e^{-2A}\dot r^2}{f^2}}}\right]
\label{cS1222h}
\ee
where now $V_3$ is the volume of the hyperbolic space instead of the volume of the 3-sphere.

The induced metric on the brane after the coordinate change \eqref{cc1} is

\be \label{cS32h}
d\hat s^2\equiv -d\tau^2+e^{2A(\tau)}R^2d\H_{3}\,,
\ee
 and the third degree algebraic equation associated with the trajectory of the brane  is
\begin{equation}
E_{H}  = e^{4 A} y\sqrt{f}[F f (y^2-1)+W_B+6\frac{U_{B}e^{-2A}}{R^2}]\,.
\label{ccS33h}
\end{equation}

Essentially, the differences in the equations in the case of spherical and hyperbolic slicings come from the relative minus sign of the curvature.

\subsection{The Mirage Cosmology}

After we have solved \eqref{cc33} or \eqref{ccS33} we can  determine the location of the brane $r(\tau)$ as a function of $\tau$ by integrating
\begin{equation}
\frac{d r}{d \tau} =\pm \sqrt{f(r)\,(y^2(r)-1)},
\label{cc5}
\end{equation}
to obtain
\be \label{cc6}
\int_{r_0}^{r(\tau)} {d r \over  \sqrt{f(r)\,(y^2(r) -1)}}  = \pm (\tau-\tau_0)
\ee
then we invert \eqref{cc6} for $r(\tau)$ in order to write the induced scale factor on the brane
\be
a(\tau) = e^{A(r(\tau))},
\label{ac1}\ee
with the corresponding induced Hubble parameter being determined by
\be
H \equiv{1\over a}{d  a \over d\tau}={ dA\over d\tau} = {dA \over dr} {dr\over d\tau} =
  \pm \frac{W(\varphi(r(\tau)))}{6} \sqrt{f(r)\,(y^2(r(\tau))-1)}
\label{ec4}\ee
The cosmological scale factor $e^{A(r)}$ is monotonically increasing
with $r$  from the IR to the UV. Moreover, $f (y^2-1)\geq 0$ always and thus the contraction or expansion of the universe is determined only by the sign of (\ref{ec4}). We conclude that as the brane moves further inside the bulk (IR) the universe contracts. Conversely as the brane heads towards the boundary of the bulk spacetime (UV) the universe expands~\cite{mirage}.

\section{Asymptotic cosmologies}\label{asym}

In this section we present the induced geometry on the probe brane when it is located close to the IR and the UV limit of the bulk geometry. This is the cosmology as seen by the observer on the brane. The parameter that defines the cosmology on the brane is the scale factor $a(\tau)$.

The UV region is reached as $r\rightarrow -\infty$ and the IR region when $r\rightarrow r_{h}$ where the black hole horizon is located.

As the brane moves towards the IR the observer perceives a contracting geometry in contrast as the brane moves in the opposite direction to the UV he/she sees an expanding universe. Since $r$ decreases monotonically as one moves from the IR to the UV, we can use the velocity $\dot{r}$ to detect the direction of the motion of the brane. If $\dot{r}<0$ it moves towards the boundary of the bulk geometry (UV) whereas if $\dot{r}>0$ it moves towards the horizon of the bulk black hole (IR).

Since we work in the IR and the UV limits our analysis is valid only approximately and thus in the expressions below we use $``\sim"$ as opposed to $``="$ when necessary.

In this note we omit the UV analysis since it is identical to the one in \cite{Amariti:2019vfv}. We just state their result for completeness. The brane geometry is  asymptotically  a dS universe, with the scale factor and the Hubble
parameter given by
\be \label{ec111}
a(\tau) \simeq a_0 e^{-\eta \tau \sqrt{{h_W\over h_U} }}\,,\quad H_{eff}\equiv {1\over a}{d  a \over d\tau}={1\over \ell}\sqrt{\frac{h_W}{h_U}}
\ee
where $\eta =\pm$, $a_0$ is set by initial conditions, $\ell$ and $h_{W,U}$ are constants governed by $\f, f$ defined in the UV region.

In the next subsections we will proceed to discuss the induced cosmology on the brane close the horizon of the bulk black hole for the cases of flat and spherical slicing.

{

\subsection{Expansion near a flat horizon\label{fho}}

To study the cosmology induced on the brane close to the black hole horizon we have to first expand the blackening factor $f(r)$, the scale factor of the bulk metric $A(r)$ and the scalar field $\f (r)$. For our purposes it is enough to keep up to order one with respect to $r-r_{h}$, where $r_{h}$ is the location of the black hole horizon. A more detail analysis can be found in appendix \ref{Afho}.

\be
f(r)\sim f_1{(r-r_h)\over \ell}
\label{e9}\ee
\be
A(r)\sim A_h+A_1{(r-r_h)\over \ell}
\label{e10}\ee
\be
\f(r)\sim \f_h+\f_1{(r-r_h)\over \ell}
\label{e11}
\ee

where $f_i,A_i,\f_i$ are dimensionless and the relations of the constants to the physical parameters of the solution can be found in the Appendix \ref{Afho}.

Substituting the expansions in \eqref{ac1}, \eqref{e26} and \eqref{c13} and keeping either zero order or up to order one in $r-r_{h}$ expansion we get

\be\label{aF}
\alpha(r_{h})=e^{2 A_{h}},\quad W_{h}=-6 A_{1},\quad   W^{\prime}_{h}=\varphi_{1},\quad F(r=r_{h})=F_{r_{h}}
\ee

Close to the horizon \eqref{cc33} becomes

\be \label{cc33h}
E  \frac{e^{-4A_{h}}\ell^{3/2}}{F_{h} f^{3/2}_{1}(r-r{h})^{3/2}}=y^3+ y (\frac{W_{h}\ell}{F_{h}f_{1}(r-r_h)}-1)
\ee

from the appendices \ref{AppCubic} and \ref{AppCubic1} we find the solution of \eqref{cc33h} to be

 \be
y\sim\frac{C}{(r-r_h)^{1/2}}
\ee

where $C>0$ is a positive constant that depends on $A_{h},A_{1},F_{h},E,\ell,f_{1}$.

Then using \eqref{cc1} we can compute the derivative of the radial direction with respect to the cosmological time $\tau$ on the brane

\be
\frac{dr}{d\tau}\sim\pm\sqrt{\frac{f_{1}}{\ell}}C
\ee
and by integrating it once we obtain the radial position of the brane from the black hole horizon $r-r_{h}$ as a function of the cosmological time $\tau$
\be
r-r_{h}=\pm\sqrt{\frac{f_{1}}{\ell}} C (\tau -\tau_{0})\,.
\ee

The corresponding Hubble parameter is
\be\label{HF}
H\sim\pm A_{1}\sqrt{\frac{f_{1}}{\ell}}C\,.
\ee

From \eqref{aF} and \eqref{HF} we conclude that the induced geometry on the brane --- as the brane approaches the horizon of the black hole--- is that of the Poincar\'{e} patch of de Sitter with scaling factor \eqref{aF} and the Hubble parameter \eqref{HF}. Depending on whether the brane moves towards the $r_{h}$ or away from $r_{h}$ the observer that lives on the brane sees a contracting $H<0$  or an expanding $H>0$ universe respectively.

Moreover, close to the black hole horizon the dS radius is $l_{dS}\sim |H|^{-1}$ and thus the temperature and the entropy associated to the dS static patch close to the bulk black hole horizon are:

\be
T_{dS}=\frac{1}{2\pi l_{dS}}=\frac{\vert A_{1}\sqrt{\frac{f_{1}}{\ell}}C\vert}{2\pi},\qquad S_{dS}=l^{2}_{dS}\pi=\frac{\pi}{ A_{1}^{2}\frac{f_{1}}{\ell}C^2}
\ee
where from appendix \ref{Afho} we find that $A_{1}=-\frac{\ell^2 V_{ h}}{2f_{1}}$ and $f_{1}$ is related to the bulk black hole temperature as $T_{BH}=\frac{f_{1}e^{A_{h}}}{4\pi}$. Thus, the de Sitter temperature (determined by H) is directly related to the bulk black hole temperature and the same holds for the $dS$ entropy with the bulk black hole entropy. Unfortunately the relation is not simple since the constant $C$ is a complicated function of the black hole temperature that comes after solving the cubic equation presented in Appendix~\ref{cubic}.

\subsection{Expansion near a spherical horizon}\label{es}

To study the cosmology induced on the brane close to the black hole horizon we have to first expand the blackening factor $f(r)$, the scale factor of the bulk metric $A(r)$ and the scalar field $\f (r)$. For our purposes it is enough to keep up to order one with respect to $r-r_{h}$, where $r_{h}$ is the location of the black hole horizon. A more detail analysis can be found in appendix \ref{Asho}.

\be
f(r)\sim f_1{(r-r_h)\over \ell}
\label{ff1}\ee
\be
A(r)\sim A_h+A_1{(r-r_h)\over \ell}
\label{ff2}\ee
\be
\f(r)\sim\f_h+\f_1{(r-r_h)\over \ell}
\label{ff3}\ee

where $f_i,A_i,\f_i$ are dimensionless and the relations of the constants to the physical parameters of the solution can be found at the appendix \ref{Asho}.

Substituting the expansions in \eqref{ac1}, \eqref{e26} and \eqref{c13} and keeping either zero order or up to order one in $r-r_{h}$ expansion we get

\be\label{aFS}
\alpha(r_{h})=e^{2 A_{h}},\quad W_{h}=-6 A_{1},\quad   W^{\prime}_{h}=\varphi_{1},\quad F(r=r_{h})= F_{h}.
\ee

Close to the horizon  \eqref{ccS33} can be written as

\be \label{ccS33ho}
E  \frac{e^{-4A_{h}}\ell^{3/2}}{F_{h} f^{3/2}_{1}(r-r{h})^{3/2}}=y^3+ y \left(\frac{-6 A_{1}\ell}{F_{h}f_{1}(r-r_h)}-6\frac{U_{h}e^{2A_{h}}}{F_{h}f_{1}(r-r_{h})R^2}-1\right)
\ee

from the appendices \ref{AppCubic} and \ref{AppCubic2} we find the solution of  \eqref{ccS33h} to be

\be
y\sim\frac{\tilde{C}}{(r-r_h)^{1/2}}
\ee

where $\tilde{C}>0$ is a positive constant that depends on $A_{h},A_{1},F_{h},E,\ell,f_{1}, R$.
Then using \eqref{cc1} we can compute the derivative of the radial direction with respect to the cosmological time $\tau$ on the brane
\be
\frac{dr}{d\tau}\sim\pm\sqrt{\frac{f_{1}}{\ell}}\tilde{C}
\ee
and by integrating it once we obtain the radial position of the brane from the black hole horizon $r-r_{h}$ as a function of the cosmological time $\tau$
\be
r-r_{h}=\pm\sqrt{\frac{f_{1}}{\ell}} \tilde{C} (\tau -\tau_{0})\,.
\ee

The corresponding Hubble parameter is
\be\label{HFS}
H\sim\pm A_{1}\sqrt{\frac{f_{1}}{\ell}}\tilde{C}\,.
\ee
From \eqref{aFS} and \eqref{HFS} we conclude that the induced geometry on the brane --- as the brane approaches the horizon of the black hole--- is that of a part of global de Sitter \footnote{Surprisingly, although in this case the slicing of the bulk geometry is spherical, we do not obtain a global dS geometry induced on the brane taking the combination of probe and IR limit.} with scaling factor \eqref{aFS} and the Hubble parameter \eqref{HFS}. Depending if the brane moves towards the $r_{h}$ or moving away from $r_{h}$ the observer that lives on the brane sees a contracting $H<0$  or an expanding $H>0$ universe respectively. At late times we can go to a flat slicing. The relation becomes $\tau_{sph} \sim \tau_{flat} $ and the metric takes the asymptotic form
\be
ds^2_{flat} \sim - d \tau_{flat}^2 + R^2 e^{2 H (\tau - \tau_0)} (dy^2 + y^2 d \Omega^2_2)
\ee
We therefore find again the relations for the temperature and entropy
\be
T_{dS}=\frac{1}{2\pi l_{dS}}=\frac{\vert A_{1}\sqrt{\frac{f_{1}}{\ell}}\tilde{C}\vert}{2\pi},\qquad S_{dS}=l^{2}_{dS}\pi R^2=\frac{\pi   R^2}{ A_{1}^{2}\frac{f_{1}}{\ell}\tilde{C}^2}
\ee
with $A_1=\frac{6\mathcal{R}^{2} - \ell^2 V(\f_{h})}{3 {f_1}}$ and  ${\mathcal{R}}\equiv {\ell\over R}e^{-A_0}$.  $f_{1}$ is related to the bulk black hole temperature as $T_{BH}=\frac{f_{1}e^{A_{h}}}{4\pi}$.

\subsection{Expansion near a hyperbolic horizon}
In the case of the hyperbolic slicing the procedure is exactly the same as in the spherical case described in subsection \ref{es}, the only difference stems from the relative minus sign in the curvature \eqref{Ricci}, thus the expansions close to the horizon \eqref{ff1} - \eqref{ff3} are the same as well as \eqref{aFS}. The expansion close to the horizon of the equation \eqref{ccS33h} is

\be \label{ccS33hoh}
E  \frac{e^{-4A_{h}}\ell^{3/2}}{F_{h} f^{3/2}_{1}(r-r{h})^{3/2}}=y^3+ y \left(\frac{-6 A_{1}\ell}{F_{h}f_{1}(r-r_h)}+6\frac{U_{h}e^{2A_{h}}}{F_{h}f_{1}(r-r_{h})R^2}-1\right)
\ee

from the appendices \ref{AppCubic} and \ref{AppCubic2} we find the solution of  \eqref{ccS33h} to be

\be
y\sim\frac{\hat{C}}{(r-r_h)^{1/2}}
\ee

where $\hat{C}>0$ is a positive constant that depends on $A_{h},A_{1},F_{h},E,\ell,f_{1}, R$.
Then using \eqref{cc1} we can compute the derivative of the radial direction with respect to the cosmological time $\tau$ on the brane
\be
\frac{dr}{d\tau}\sim\pm\sqrt{\frac{f_{1}}{\ell}}\hat{C}
\ee
and by integrating it once we obtain the radial position of the brane from the black hole horizon $r-r_{h}$ as a function of the cosmological time $\tau$
\be
r-r_{h}=\pm\sqrt{\frac{f_{1}}{\ell}} \hat{C} (\tau -\tau_{0})\,.
\ee

The corresponding Hubble parameter is
\be\label{HFS}
H\sim\pm A_{1}\sqrt{\frac{f_{1}}{\ell}}\hat{C}\,.
\ee
From \eqref{aFS} and \eqref{HFS} we conclude that the induced geometry on the brane --- as the brane approaches the horizon of the black hole--- is that of a part of de Sitter with hyperbolic slicing \footnote{Surprisingly, although in this case the slicing of the bulk geometry is hyperbolic, we do not obtain a hyperbolic dS geometry induced on the brane taking the combination of probe and IR limit.} with scaling factor \eqref{aFS} and the Hubble parameter \eqref{HFS}. Depending if the brane moves towards the $r_{h}$ or moving away from $r_{h}$ the observer that lives on the brane sees a contracting $H<0$  or an expanding $H>0$ universe respectively. At late times we can go to a flat slicing. The relation becomes $\tau_{hyp} \sim \tau_{flat} $ and the metric takes the asymptotic form
\be
ds^2_{flat} \sim - d \tau_{flat}^2 + R^2 e^{2 H (\tau - \tau_0)} (dy^2 + y^2 d
\H^2_2)
\ee
We therefore find again the relations for the temperature and entropy
\be
T_{dS}=\frac{1}{2\pi l_{dS}}=\frac{\vert A_{1}\sqrt{\frac{f_{1}}{\ell}}\hat{C}\vert}{2\pi},\qquad S_{dS}=l^{2}_{dS}\pi R^2=\frac{\pi   R^2}{ A_{1}^{2}\frac{f_{1}}{\ell}\hat{C}^2}
\ee
with $A_1=\frac{-6\mathcal{R}^{2} - \ell^2 V(\f_{h})}{3 {f_1}}$ and  ${\mathcal{R}}\equiv {\ell\over R}e^{-A_0}$.  $f_{1}$ is related to the bulk black hole temperature as $T_{BH}=\frac{f_{1}e^{A_{h}}}{4\pi}$.

We now conclude this section, by summarizing our results. Remarkably, for all the slicings (flat, spherical and hyperbolic) of the bulk black hole geometry, we find that the cosmologies induced on the probe brane both in the UV and in the IR are de-Sitter spacetimes, with different scaling factors and Hubble parameters. Thus we conclude that the cosmological models constructed in our paper interpolate between two de Sitter geometries. This is in contrast with the cases studied in \cite{Amariti:2019vfv}, where the probe brane acquires a big bang/crunch singularity when its location is close to the bulk's IR. In particular in our regular horizon bulk geometries, the big bang/crunch singularity is encountered only when the brane passes through the horizon and reaches the bulk black hole singularity.

%%%%%%%%%%%%%%%%%%%%%%%%%%%%%%%%%
\section*{Acknowledgements}\label{ACKNOWL}
\addcontentsline{toc}{section}{Acknowledgements}

We wish to thank Elias Kiritsis for useful suggestions. We also thank Francesco Nitti for useful discussions and comments. We happily acknowledge the hospitality provided by APC Paris, during the initial stages of this work.

\noindent This work is supported in part by the Advanced ERC grant SM-GRAV, No 669288.

%%%%%%%%%%%%%%%%%%%%%%%%%%%%%%%%%%%%%%%%%%%%%%%%%%%%%
\newpage
\appendix
\renewcommand{\theequation}{\thesection.\arabic{equation}}
\addcontentsline{toc}{section}{Appendix\label{app}}
\section*{Appendices}

\section{Black-hole ansatz with a flat slicing\label{flat}}

{ In this appendix we present the equations of motion for the Einstein Dilaton action when the ansatz of the metric is a black hole with flat slicing as the one in  \eqref{e1}. For completeness we rewrite here the ansatz for the metric
\be
ds^2={dr^2\over f(r)}+e^{2A(r)}\left[-f(r)dt^2+dx_{i}dx^{i}\right].
\label{e11}\ee

The equations of motion for the action \eqref{A2} are}

\begin{subequations}\label{e2}
\begin{align}
	&2(d-1)\ddot{A}(r)+\dot{\f}^2(r)=0,\label{e2a}\\
	&\ddot{f}(r)+d\dot{f}(r)\dot{A}(r)=0\label{e2b}\\ &(d-1)\dot{A}(r)\dot{f}(r)+f(r)\left[d(d-1)\dot{A}^2(r)-\frac{\dot{\f}^2}{2}\right]+V(\f)=0.\label{e2c}
\end{align}
\end{subequations}
\be
\ddot\f(r)+\left(d\dot A(r)+{\dot f(r)\over f(r)}\right)\dot\f(r)-{V'\over f(r)}=0
\label{e2d}\ee
where $\dot{~}$ is the derivative with respect to the radial coordinate.

We integrate \eqref{e2b} once to obtain
\be
 \dot f=C~e^{-dA}
\label{e2e}\ee
where $C$ is an integration constant.

From \eqref{e2a} we can express \eqref{e2c} as
\be
(d-1)\left[\dot f\dot A+f(d\dot A^2+\ddot A)\right]+V=0
\label{e3}\ee
or
\be
(d-1){d\over dr}\left(f\dot A~e^{dA}\right)=-V~e^{dA}
\label{e4}\ee
Moreover, \eqref{e2d} can be written as
\be
{d\over dr}\left(f\dot \f~e^{dA}\right)=V'~e^{dA}
\label{e5}\ee
where $\prime$ is a derivative with respect to the scalar field.
For $f(r)=1$, and $A(r)=r/\ell$, \eqref{e1} is an AdS metric, with AdS radius $\ell$. For $f(r)=-1$ the metric is dS with radius $\ell$
\be
A(r)=r H\sp  H={1\over \ell}~~\;.
\ee

\subsection{The superpotential formalism}\label{ASF1}

In this subsection we present the superpotential formalism. We begin by defining the superpotential $W$ as usual

\be \label{e26}
W(\f) \equiv -2(d-1)\dot A (r)\,,\quad \dot\f=W'\;.
\ee
By using
\be
\pa_r=W'(\f)\pa_{\f}
\label{w29}\ee
the equation\eqref{e2b} becomes a function of $\varphi$
\be
W'f''+\left(W''-{d\over 2(d-1)}W\right)f'=0
\label{e28}\ee
which we integrate once to obtain
\be\label{e28_a}
f'={1\over W'}\exp\left[{d\over 2(d-1)}\int d\f {W\over W'}\right]
\ee

Then using  \eqref{e26} and \ref{e2c}

\be\label{e29}
\left(\frac{d}{4 (d-1)}W^2(\f)-\frac{(W'(\f))^2}{2}\right)f(u)-\frac{\dot{f}(u)}{2} W(\f)+V(\f)=0.
\ee
or
\be
\left(\frac{d }{4 (d-1)}W^2(\f)-\frac{(W'(\f))^2}{2}\right)f(\f)-{W'\over 2}f'(\f) W(\f)+V(\f)=0.
\label{e30}\ee

Then we rewrite \eqref{e2b} as
\be
f'={C\over W'}e^{-dA}
\label{w29a}\ee
such that \eqref{e30} becomes
\be
{\bigg(\frac{d }{4 (d-1)}W^2(\f)-\frac{(W'(\f))^2}{2}\bigg)f(\f)-{C\over 2} W(\f)e^{-dA}+V(\f)=0.}
\label{e31}\ee

Moreover, the integration of \eqref{e2b}, gives

\be
\label{e32}f(r)=\int_{-\infty}^{r} dr'\exp \left[ \frac{d}{2(d-1)} \int_{\f_*}^{\f(r')}d\f' \frac{W(\f')}{W'(\f')} \right]+1,
\ee
where $f(r) \rightarrow 1$ as $r \rightarrow -\infty$, and $\f_*$ is an integration constant.

A more detailed analysis can be found in \cite{Kiritsis:2019wyk}.

\subsection{Expansion near a flat horizon\label{Afho}}

In this subsection we present the expansion of the blackening factor $f(r)$, the scalar field $\f(r)$ and of the scale factor $A(r)$ close to the black hole horizon at $r=r_h$,

\be
f(r)=f_1{(r-r_h)\over \ell}+{f_2\over 2}{(r-r_h)^2\over \ell^2}+{f_3\over 3!}{(r-r_h)^3\over \ell^3}+{\cal O}\left((r-r_h)^4\right)
\label{e9}\ee
\be
A(r)=A_h+A_1{(r-r_h)\over \ell}+{A_2\over 2}{(r-r_h)^2\over \ell^2}+{A_3\over 3!}{(r-r_h)^3\over \ell^3}+{\cal O}\left((r-r_h)^4\right)
\label{e10}\ee
\be
\f(r)=\f_h+\f_1{(r-r_h)\over \ell}+{\f_2\over 2}{(r-r_h)^2\over \ell^2}+{\f_3\over 3!}{(r-r_h)^3\over \ell^3}+{\cal O}\left((r-r_h)^4\right)
\label{e11}\ee
We have defined the coefficients $f_i,A_i,\f_i$ such that they are dimensionless.

Substituting the expansions above into \eqref{e2a}-\eqref{e2d} we obtain
\be
(d-1)A_1f_1+\ell^2~V_h=0\sp f_1\f_1-\ell^2~V'_h=0
\label{e12}\ee
and
\be
2f_1\f_2=\f_1~\ell^2~V''_h\sp A_2=-{\f_1^2\over 2(d-1)}\sp f_2=-{d}f_1A_1
\label{e13}\ee
with the following definitions
\be
V_h\equiv V(\f_h)\sp V'_h\equiv V'(\f_h)\,.
\label{e14}\ee
In order for the horizon to be regular, $f_1\not =0$ which implies
\be
A_1=-{\ell^2~V_h\over (d-1)f_1}\sp \phi_1={\ell^2~V'_h\over f_1}
\label{e15}\ee
\be
\f_2={\ell^4~V'_hV''_h\over 2f_1^2}\sp A_2=-{\ell^4~V_h'^2\over 2(d-1)f_1^2}\sp f_2={d\over (d-1)}\ell^2~V_h
\label{e16}\ee
\be
\f_3=\ell^6~{V'_h(d(V'_h)^2-dV_hV''_h+(d-1)(V''_h)^2+2(d-1)V'_hV'''_h)\over 6(d-1)f_1^3}
\label{e17}\ee
\be
A_3=-\ell^6~{(V'_h)^2V''_h\over 2(d-1)f_1^3}\sp f_3=\ell^4~{d(2dV_h^2+(d-1)(V'_h)^2)\over 2(d-1)^2f_1}
\label{e18}\ee

The above is a solution even in the case where the potential vanishes at the horizon $V(\f_h)=0$. The AdS-Schwarzschild black hole is obtained in the case that the scalar potential has an extremum on the horizon $V'(\f_h)=0$. In this case $\f=constant$ and $A$ is linear in $r$.

In the case that we expand around $r=r_{h}$, the solutions to the equations of motion are governed only by three parameters. This can be seen from the equation \eqref{e15}-\eqref{e18} combined with \eqref{e12} which relates $f_{1}$ with $\f_{1}$. This has the consequence that when $V_{h}=0$ then from \eqref{e15} $A_1\sim \dot A=0$ at $r=r_{h}$.

In the case that $V'(\f_h)=0$ \eqref{e15} -\eqref{e18} become

\be
A_1=-{V(\f_h)\over (d-1)f_1}\sp A_2=A_3=\cdots=0\sp \f_1=\f_2=\cdots=0
\label{w25}\ee
\be
f_2={d\over d-1}V(\f_h)\sp f_3={d^2\over (d-1)^2}{V^2(\f_h)\over f_1}
\label{w26}\ee
the blackening factor can be resummed as
\be
f(r)={f_1\over dA_1}\left[1-e^{-dA_1 r}\right]\, ,
\label{w27}\ee
where $\f$ is constant. From this equation we can observe that the location of the horizon can be at any point.

\section{Black hole ansatz with a spherical slicing}

In this appendix we present the equations of motion for the Einstein Dilaton action when the ansatz of the metric is a black hole with spherical slicing as the one in  \eqref{m1}. For completeness we rewrite here the ansatz for the metric

\be
ds^2={dr^2\over f(r)}+e^{2A(r)}\left[-f(r)dt^2+R^2~d\Omega_{d-1}^2\right].
\label{f22}\ee
with $d\Omega^2_{d-1}$ we denote the $(d-1)$-dimensional sphere metric with radius one and $R$ is the radius of the sphere. We are interested in the case $d=4$, but our method applies for cases $d>2$.
The equations of motion are
\begin{subequations}\label{f23}
\begin{align}
	&6\ddot{A}(r)+\dot{\f}^2(r)=0,\label{f23a}\\
	&\ddot{f}(r)+4\dot{f}(r)\dot{A}(r)+{4\over R^2} e^{-2 A(r)}=0\label{f23b}\\
	 &3\dot{A}(r)\dot{f}(r)+f(r)\left[12\dot{A}^2(r)-\frac{\dot{\f}^2}{2}\right]+V(\f)-{6\over R^2} e^{-2 A(r)} =0.\label{f23c}
\end{align}
\end{subequations}
\be
\ddot\f(r)+\left(4\dot A(r)+{\dot f(r)\over f(r)}\right)\dot\f(r)-{V'(\f)\over f(r)}=0
\label{f23d}\ee
We rewrite \eqref{f23b} as
\be
\left(\dot f~e^{4A}\right)\dot{\phantom{I}} =-{4\over R^2}e^{2A}
\label{f13}\ee
and  \eqref{f23c} as
\be
3\left(f\dot A~e^{4A}\right)\dot{\phantom{I}} +e^{4A}~V-{6\over R^2}e^{2A}=0
\label{f12}\ee
We then integrate \eqref{f23b} to obtain
\be
\dot f(r)=Ce^{-4A(r)}-{4e^{-4A(r)}\over R^2}\int_{r_0}^{r}dr'~e^{2A(r')}
\label{w52}\ee

A more detailed analysis can be found in \cite{Kiritsis:2019wyk}.

\subsection{The superpotential formalism}\label{BSF1}

As in all the previous cases, we introduce a superpotential by defining

\be
W(\f) \equiv - 6 \dot{A}(r), \ \dot{\f}=W'(\f).
\label{w53}\ee
We notice that \eqref{f23a} is satisfied automatically and \eqref{f23b} becomes

\be \label{f10_1}
W' \bigg[ W' f''+\bigg( W''-\frac{2}{3} W\bigg)f'\bigg]+4T(\f)
=0,
\ee
where
\be
T(\f)\equiv \frac{1}{R^2} e^{-2 A(\f)}\sp T'={W\over 3W'}T
\label{w54a}\ee
and
\be
A(\f)=-\frac{1}{6} \int_{\f^*}^\f \frac{W(\f')}{W'(\f')} d \f'.
\label{w54}\ee
Moreover \eqref{f23c} is

\be
\bigg( \frac{1}{3} W^2(\f)-\frac{(W'(\f))^2}{2}\bigg) f(\f)-\frac{1}{2} W'(\f)W(\f)f'(\f)+V(\f)-6T(\f)=0.
\label{w55}\ee

We integrate \eqref{f10_1} to get

{
\be
f'_S (\f)=-4f'_F(\f)\int_{\f^*}^\f \frac{T(\f')}{f'_F (\f') W'(\f')^2} d \f',
\label{w56}\ee

with $F$ referring to flat slicing and $S$ to the spherical one.

\subsection{Solutions near a spherical horizon\label{Asho}}

In this subsection we present the expansion of the blackening factor $f(r)$, the scalar field $\f(r)$ and of the scale factor $A(r)$ close to the black hole horizon at $r=r_h$,

\be
f(r)=f_1{(r-r_h)\over \ell}+{f_2\over 2}{(r-r_h)^2\over \ell^2}+{f_3\over 3!}{(r-r_h)^3\over \ell^3}+{\cal O}\left((r-r_h)^4\right)
\label{f1}\ee
\be
A(r)=A_h+A_1{(r-r_h)\over \ell}+{A_2\over 2}{(r-r_h)^2\over \ell^2}+{A_3\over 3!}{(r-r_h)^3\over \ell^3}+{\cal O}\left((r-r_h)^4\right)
\label{f2}\ee
\be
\f(r)=\f_h+\f_1{(r-r_h)\over \ell}+{\f_2\over 2}{(r-r_h)^2\over \ell^2}+{\f_3\over 3!}{(r-r_h)^3\over \ell^3}+{\cal O}\left((r-r_h)^4\right)
\label{f3}\ee
\def\cR{{\cal R}}
We have defined the coefficients $f_i,A_i,\f_i$ such that they are dimensionless.

Substituting the expansions above into \eqref{f23}-\eqref{f23c} we obtain
\be
\f_1={\ell^2 V'(\f_h)\over f_1}\sp {A_1}=\frac{{6\cR^2} - \ell^2 V({\f_h})}{3 {f_1}}
\label{f4}\ee
\be
{f_2}= \frac{4}{3} \left(\ell^2V({\f_h})-9\cR^2\right)\sp {A_2}= -\frac{\ell^4V'({\f_h})^2}{6{f_1}^2}
\label{f5}  \ee
  \be
    {\f_2}= \frac{ \ell^2V'({\f_h}) \left(\ell^2V''({\f_h})+{4\cR^2}\right)}{2 {f_1}^2}
 \label{f6}  \ee
   \be
   {f_3}= \frac{2 \left(-132 \cR^2 \ell^2V({\f_h})+504\cR^4
+3 \ell^4V'({\f_h})^2+8
   \ell^4V({\f_h})^2\right)}{9{f_1}}
\label{f7}   \ee
\be
{A_3}= -\frac{ \ell^4 V'({\f_h})^2 \left( \ell^2V''({\f_h})+{4\cR^2}\right)}{6{f_1}^3 }
  \label{f8} \ee
   \be
   {\f_3}=4\cR^2{\ell^4 V'({\f_h})
   V''({\f_h})\over f_1^3}+8\cR^4 {\ell^2V'({\f_h})\over f_1^3}+
\label{f9}   \ee
   $$+\ell^6
   \frac{4
   V'({\f_h})^3+6 V^{(3)}({\f_h}) V'({\f_h})^2+3
    V'({\f_h}) V''({\f_h})^2-4V({\f_h})
   V'({\f_h}) V''({\f_h})}{18 {f_1}^3}
   $$
   where
   \be
   {\cR}\equiv {\ell\over R}e^{-A_0}
 \label{f10}  \ee

\section{Solving the cubic equation}\label{cubic}
\label{AppCubic}
In this appendix, we present the possible solutions of \eqref{cc33}, \eqref{ccS33}. For a more detailed analysis one can look at the appendix D of \cite{Amariti:2019vfv}.

In general a cubic equation of the form
\be
y^3 + b y + c = 0,
\quad
\text{with}
\quad
b\,, c \in\mathcal{C}
\ee

has the general solution
\be
y_{n}=e^{in\frac{2\pi}{3}}\left[\frac{1}{2}\left(-c+\sqrt{\frac{\Delta_{3}}{27}}\right)\right]^{1/3}+e^{-in\frac{2\pi}{3}}\left[\frac{1}{2}\left(-c-\sqrt{\frac{\Delta_{3}}{27}}\right)\right]^{1/3},\quad n=\left(0,1,2\right).
\ee
where
\be
\Delta_3 = -(4 b^3+27 c^2)
\ee

We summarize the various possibilities taking into account the condition $y>1$. The details can be found in appendix D of \cite{Amariti:2019vfv}
\begin{center}
\begin{tabular} {|c|c|}
\hline
& Number of solutions with $y>1$ \\
\hline
$\Delta_3<0$ & One, if $b<-(c+1)$         \\
\hline
$\Delta_3=0$ &
Two (coincident), if  $c>2$
\\
\hline
$\Delta_3=0$ &One, if $c<-\frac{1}{4}$   \\
\hline
$\Delta_3>0$ & One, if  $b<-(c+1)$      \\
\hline
$\Delta_3>0$ & Two, if $c>2$ and $b>-(c+1)$   \\
\hline
\end{tabular}
\end{center}

\subsection{Solution close to the flat horizon}\label{AppCubic1}

We can write equation \eqref{cc33} as

\begin{equation}
E  \frac{e^{-4A_{h}}}{F f^{3/2}} = y^3+y (\frac{W}{F f}-1)
\label{ccc33}
\end{equation}
with
\be
b=\frac{W}{F f}-1\,\quad c= -E  \frac{e^{-4A_{h}}}{F f^{3/2}}
\ee

Close to the horizon it becomes
\be
E  \frac{e^{-4A_{h}}\ell^{3/2}}{F_{h} f^{3/2}_{1}(r-r{h})^{3/2}}=y^3+ y (\frac{W_{h}\ell}{F_{h}f_{1}(r-r_h)}-1)
\ee

with
\be
b=\frac{W_{h}\ell}{F_{h}f_{1}(r-r_h)}\,\quad c= -E \frac{e^{-4A_{h}}\ell^{3/2}}{F_{h} f^{3/2}_{1}(r-r{h})^{3/2}}
\ee

\be
\Delta_{3h}=\frac{-4\times 6^3 A_{1}^3 \ell^3+27 F_{h}E^2 e^{-8 A_{h}}\ell^3}{4\times 27 F_{h}^3 f_{1}^{3}(r-r_h)^3}
\ee
Depending on the sign of $\Delta_{3}$ we have different solution given in \ref{AppCubic}, but all have the form
\be
y\sim\frac{C}{(r-r_h)^{1/2}}
\ee
where $C>0$ a positive  constant that depends on $A_{h},A_{1},F_{h},E,\ell,f_{1}$.

\subsection{Solution close to the spherical and to the hyperbolic horizon}\label{AppCubic2}

We can write \eqref{ccS33} and \eqref{ccS33h} as

\begin{equation}
E  \frac{e^{-4A}}{F f^{3/2}} = y^3+y (\frac{W\mp\frac{6 U_{B}e^{2A}}{R^2}}{F f}-1)
\label{cccS33}
\end{equation}
where the upper sign is for the spherical case and the lower for the hyperbolic one, with

\be
b=\frac{W}{F f}\mp 6\frac{e^{2A}U_{B}}{Ff R^2}-1\,\quad c= -E  \frac{e^{-4A}}{F f^{3/2}}
\ee
and

Close to the horizon they become

\be
E  \frac{e^{-4A_{h}}\ell^{3/2}}{F_{h} f^{3/2}_{1}(r-r_{h})^{3/2}}=y^3+ y \left(\frac{-6 A_{1}\ell}{F_{h}f_{1}(r-r_h)}\mp 6\frac{U_{h}e^{2A_{h}}}{F_{h}f_{1}(r-r_{h})R^2}-1\right)
\ee

and

\be
b=\frac{-6 A_{1}\ell}{F_{h}f_{1}(r-r_h)}\mp 6\frac{U_{h}e^{2A_{h}}}{F_{h}f_{1}(r-r_{h})R^2}\,\quad c= -E \frac{e^{-4A_{h}}\ell^{3/2}}{F_{h} f^{3/2}_{1}(r-r{h})^{3/2}}
\ee

\be
\Delta_{3h}=\ell^3\frac{-4\times 6^3 (A_{1}\mp U_{h} e^{2A_h}/R^2)^3 +27 F_{h}E^2 e^{-8 A_{h}}}{4\times 27 F_{h}^3 f_{1}^{3}(r-r_h)^3}
\ee

Depending on the sign of $\Delta_{3}$ we have different solutions given in appendix \ref{AppCubic}, but all of them have the form

\be
y_s\sim\frac{\tilde{C}}{(r-r_h)^{1/2}}\,, \quad y_h\sim\frac{\hat{C}}{(r-r_h)^{1/2}}
\ee

where $\tilde{C},\,\hat{C}>0$ positive constants that depends on $A_{h},A_{1},F_{h},E,\ell,f_{1},R$.

\section{Embedding of a Black hole with hyperbolic slicing}\label{hyp}

We now examine the case of hyperbolic slicing. This corresponds to the ansatz
\be
ds^2={dr^2\over f(r)}+e^{2A(r)}\left[-f(r)dt^2+R^2 d\H_{d-1}^{2}\right].
\label{mm1}\ee
where $d\H_{d-1}$ is the metric of the transverse sphere. In the case of five bulk dimensions the explicit metric is
\be
d\H_{3}^{2}=d\psi^2+\sinh^2\psi\left(d\theta^{2}+\sin^{2}\theta\,d\phi^2\right)
\ee
The Ricci scalar induced on the brane is
\be
R^{\gamma}=-\frac{6 e^{-2 A}}{R^2}=-\frac{(d-2)(d-1) e^{-2 A}}{R^2}
\ee

The Einstein tensor on the brane is
\be
G^{\gamma}_{\mu\nu}=\begin{pmatrix}
\frac{-3 f(r)}{R^2} & 0 & 0 & 0\\
0 & 1 & 0 & 0 \\
0 & 0 & \sinh^2 \psi  & 0\\
0 & 0 & 0 & \sinh^2 \psi \sin^2 \theta
\end{pmatrix}
\ee

We substitute the above at the equation \eqref{FE5} and we find the following two relations

\be\label{eE1}
\left[K_{tt}-\gamma_{tt}K\right]_{UV}^{IR}=(d-1)e^{2A}f(r)^{3/2} \left[ \dot{A} \right]_{UV}^{IR}=-\frac{1}{2} W_{B}e^{2A}f(r) +U_{B}(\varphi)(d-1) \frac{f(r)}{R^2}
\ee

\be\label{e2}
\left[K_{ij}-\gamma_{ij}K\right]_{UV}^{IR}=-e^{2A}R^2\chi_{ij}(\Omega_{d-1})\sqrt{f(r)}\left[(d-1)\, \dot{A}+\frac{1}{2}\frac{\dot{f(r)}}{f(r)}\right]_{UV}^{IR}
\ee
$$ =-\frac{1}{2} W_{B}e^{2A} R^2 \chi_{ij}(\Omega_{d-1})+U_{B}(\varphi)G_{ij}^{\gamma}=-\frac{1}{2} W_{B}e^{2A} R^2 \chi_{ij}(\Omega_{d-1})-U_{B}(\varphi)\chi_{ij}(\Omega_{d-1})$$

For \eqref{eE1} and \eqref{e2} to be consistent the following equation should then be satisfied

\be \label{conn2}
\frac{U_{B}(d-2)}{R^2}=-\frac{e^{2A}}{2} \left[ \frac{\dot{f(r)}}{\sqrt{f(r)}} \right]_{UV}^{IR}
\ee
Equation \eqref{FE6} then becomes

\be\label{s2}
n^{\alpha}\partial_{a}\varphi\vert_{UV}^{IR}=\sqrt{f(r)} \left[ \partial_{r}\varphi \right]_{UV}^{IR}=\frac{dW_{B}}{d\varphi}+\frac{e^{-2A}(d-2)(d-1)}{R^2}\frac{d U_{B}}{d\varphi}
\ee

We notice that the only difference comparatively to the spherical slicing is in \eqref{s2} where there is a minus sign in front of the second term due to the opposite sign in the curvature. The equation \eqref{conn2} remains the same since the explicit form of the transverse metric drops out from the equations.

%%%%%%%%%%%%%%%%%%%%%%%%%%%%%%%%%%%%%%
\addcontentsline{toc}{section}{References}
%######################################################################################################
 
\end{document}